\documentclass[journal=mamobx,manuscript=article]{achemso}
\setkeys{acs}{articletitle = true}

\usepackage{chemformula} 
\usepackage[T1]{fontenc} 
\usepackage{color} 
\usepackage{amsmath} 

\author{Nicol\'as A. Garc\'ia}
\email{garciana@ill.fr}
\author{Jean-Louis Barrat}
\affiliation{Institut Laue-Langevin, 71 Avenue des Martyrs, 38042 Grenoble, France}
\alsoaffiliation{Univ. Grenoble Alpes, CNRS, Liphy, 140 Rue de la Physique, 
38402 Saint-Martin-d'H\`eres, France}
\title[]{Entanglement reduction induced by geometrical confinement in polymer thin films}

\abbreviations{IR,NMR,UV}
\keywords{American Chemical Society, \LaTeX}

\begin{document}

%
%
%
%
%

\begin{abstract}
We report simulation results on melts of entangled linear polymers confined in a 
free-standing thin film. We study how the geometric constraints imposed by the 
confinement alter the entanglement state of the system compared to the 
equivalent bulk system using various observables. We find that the confinement 
compresses the chain conformation uniaxially, decreasing the volume pervaded by 
the chain, which in turn reduces the number of the accessible inter-chain 
contact that could lead to entanglements. This local and non-uniform effect 
depends on the position of the chain within the film. We also test a recently 
presented theory that predicts how the number of entanglements decreases with 
geometrical confinement.
\end{abstract}

\section{Introduction}
Viscoelastic properties of polymer in melts or concentrated solutions depend 
strongly on the molecular weight of the polymer chains. The main effect of 
increasing molecular weight is the apparition of topological constraints between 
the chains called entanglements. These constraints are a universal aspect of 
polymer physics and arise in any flexible polymer if the chain is sufficiently 
long and the concentration is high enough. Under these conditions, the effect of 
entanglements becomes so relevant that system dramatically changes their 
physical properties such as viscosity, diffusion, rheological and mechanical 
behavior.

Nowadays, the most extended and successful theory regarding entangled polymer 
dynamics is the \textit{Tube model} presented by Doi and Edwards 
\cite{Doi_Edwars_1988} and extended as \textit{Reptation model} by de Gennes 
\cite{DeGennes1971,DeGennes1976} which provided a framework for understanding 
many aspects of the underlying polymer physic in both regimes: melt and 
solution. The theory averages the collective effect of all surroundings chains 
over a given strand to a tubelike-region of confinement whose central axis is 
one segment of the called \textit{primitive path} ($PP$). In this approach, the 
primitive path is an essential theoretical concept introduced by Edwards 
\cite{Edwards1965,Edwards1967} and defined as the shortest path connecting the 
two ends of the chain preserving its topology. As a result of this confinement 
in a virtual tube, the strand moves back and forth performing a 
\textit{slithering motion} inside the tube (reptates). Despite its conceptual 
simplicity this theory proved to be a powerful tool to understand polymer 
dynamics and their quantitative predictions fit quite well with experimental 
results.

Polymer thin films have numerous applications (e.g., coatings, dielectrics, 
adhesives, lubricants\cite{Rayss1993,Zhang1999,Marencic2010}), but are also of 
fundamental interest. Thin films below a certain thickness induce geometrical 
confinement so that the polymeric material exhibits unusual physical properties 
compared to its bulk behavior\cite{Tsui2001,Rathfon2011}. Viscoelastic 
properties are not the exception and are affected below of a certain confinement 
strength. This is clearly reported in experiments\cite{Campise2017,Aoki2008}; 
however, the precise link between these modifications and the changes in the 
topological structure, or entanglement network, are not fully understood yet. 
Indeed, the manner in which the entanglement network is modified under 
confinement is a subject of current interest. 

As entanglements are not directly observable via experiments, numerical 
simulations are an essential tool to study their nature. Since entangled chains 
have very long relaxation times, classical Molecular Dynamics is quite limited 
for such studies. Recently new coarse-graining techniques were introduced to 
simulate entanglements such as \textit{slip-springs} or \textit{slip-links} 
which introduce a temporary attractive force between nearby beads, imitating 
entanglement 
effects\cite{Chappa2012,Delbiondo2013,Masnada2013,Ramirez2013,Ramirez2015, 
Ramirez2017}. However, in such studies, the effect of heterogeneity or 
confinement on the slip link (entanglement) density has to be specified somewhat 
arbitrarily, so that it becomes essential to inform such techniques using a more 
microscopic approach, numerical or theoretical. 

Recently, a step in this direction was performed by extending the principle of 
conformational transfer of Silberberg to predict the entanglement reduction in 
thin films or cylinders as a function of the aspect ratio between the film 
thickness (or cylinder radius) and the end-to-end chain 
distance\cite{Sussman2014,Sussman2016}. The predictions of the theory were 
tested using molecular dynamics simulations, however in weakly entangled 
systems.

In this paper, we extend this analysis to more strongly entangled systems, using 
a technique that uses ultrasoft potentials to speed up the simulation 
\cite{Korolkovas2016}. Our primary aim is to unveil how, in a thin film built 
with linear polymers, the geometrical confinement acts as an external field that 
modifies the entanglement state of the system, and what are its most relevant 
consequences.
{The article is organized as follow: In the next section, we 
describe the simulation model and the methods and protocols used in our study. 
Then, the section \textit{``Results and discussion''} presents the results,
with five subsections addressing different aspects of the confinement 
effect. The main conclusions are summarized in the last section.}

\section{Model and methods}
\label{sec:1}
\hspace{\parindent}
The model is based on a new original approach to simulate entangled of polymer 
in melt or concentrated solution condition, reported in an earlier work 
\cite{Korolkovas2016}, and which was recently successfully used to study polymer 
brushes under shear flow\cite{Korolkovas2017}.

The main idea is to use a pseudo-continuous model of a polymer solution, 
consisting of long chains interacting through a soft potential field. The motion 
is then resolved using Brownian dynamics with large time steps.

The motion of $C$ chains in dense conditions, each described by a continuous 
curve $\mathbf{R}_{c}(t, s)$, with variables $t$ for time and $s \in (0,1)$ as 
the monomer index, is solved numerically. The continuous backbone $s$, uses a 
finite number of discrete points $j=1,2,...,J$ to, generally, oversample the 
chains. Choosing $J=N$ the chain is reduced to the standard bead-spring model, 
which for this soft-potential has gaps that may allow chains to cross each 
other. This is a novelty aspect on this coarse-graining, where the crosses are 
avoided oversampling enough the chains to suppress the gaps along the backbone 
effectively. In this work, we found that $J=4N$ is sufficient to describe the 
chains in all our simulations well.

Every chain has $N$ degrees of freedom that correspond to the usual Rouse modes 
(or alternatively to $N$ beads through the usual Rouse transformation, see Ref. 
\citenum{Korolkovas2016}), and follows the stochastic first order equation of 
motion:
\begin{eqnarray}
\zeta \frac{\partial \mathbf{R}_{c}(t, s)}{\partial t} = F_{s} - N \nabla V_{c} + \sqrt{2k_B T\zeta}\mathbf{W}_{c}(t, s)
\label{Eq:Motion}
\end{eqnarray}
here $\zeta= N\zeta_0$ 
is the friction coefficient of the chain center of mass. 

The strength of the thermal noise is 
modeled by a Wiener process $<\mathbf{W}_{c}(t, s) \mathbf{W}_{c'}(t', 
s')>=\delta_{cc'}\delta(t-t')\delta(s-s')$.

In Eq. \ref{Eq:Motion} $F_s$ models the bonded interaction (bead-spring):
\begin{eqnarray}
F_{s} = \left(\frac{3 k_b T}{Nb^2}\right) \frac{\partial^2 \mathbf{R}_{c}(t, s)}{\partial s^2}.
\end{eqnarray}
{where} $Nb^2$ is the {mean square} 
end-to-end distance of a free chain, and can be combined with other parameters 
to define the microscopic unit of time, $\tau= \zeta_0 b^2/k_BT$. 

\noindent $V_{c}$ describes the nonbonded interactions between chains:
\begin{eqnarray}
V_{c} = \sum_{c'=1}^{C} \int_0^1 \Phi[\mathbf{R}_{c}(t, s)-\mathbf{R}_{c'}(t, s')] ds'
\end{eqnarray}
Here, we propose as $\Phi(r)$ a soft potential model through a combination of 
Gaussian functions, that takes into account both relevance interaction; excluded 
volume and attractive force:
\begin{eqnarray}
 \Phi(r) = \left(\frac{N}{J}\right) k_B T \left[(w+1)e^{-r^2/2\lambda^2} - we^{-r^2/4\lambda^2} \right]
 \label{Eq:SoftPotential}
\end{eqnarray}
\noindent where $w\geq0$ is a parameter to control the relative weight of the 
attractive part. At this point, it is important to remark that in potentials 
such as the one proposed in Eq. \ref{Eq:SoftPotential} a problem of thermodynamic 
stability may arise, so the selection of the value for $w$ is not a trivial 
question. In fact is well known\cite{Louis2000,Heyes2008,Heyes2010} that when 
interacting bodies without an infinitely repulsive core (i.e., finite force 
value at zero separation) also interact through attractive forces, if the 
attraction is too strong the weak short-range repulsion may not be sufficient to 
prevent a ``collapse'' of the system allowing all particles eventually to 
overlap in a finite region of space: the thermodynamic catastrophe occurs. 
Well-defined criteria to ensure thermodynamic stability were derived by Fisher 
and Ruelle\cite{FisherRuelle,Ruelle1999}. In this work, we determined a 
thermodynamic safety interval of values for $w$ applying these criteria 
following a straightforward approach proposed in Ref. \citenum{Heyes2007}. We found 
that the stability condition is $w \geq (2^{3/2}-1)^{-1}$ for this parameter, 
and we use the value $w=0.5$ for all the simulation reported here. More details 
are given in Appendix I.

A central point of this coarse-grained description is the use of an approximate 
but high-speed method of evaluating the interparticle forces. The technique 
involves splitting the force into two terms and evaluate them on a staggered 
grid. The first term takes into account short-range interactions and the second 
one the long-range contributions. The short-range part is calculated through 
linearization of the gaussian force, and the long-range part using a convolution 
between the density field and the potential in Fourier space, where the periodic 
boundary conditions are incorporated naturally.

The matrix-matched nature of this problem allows the implementation of a 
transparent parallelization into the GPU paradigm, which enabled us to take 
advantage of these high-performance devices. Hence, the simulation code was 
programmed in CUDA with an optimized implementation of the available memories 
and all simulations reported in this manuscript were run in two GPU cards Nvidia 
Quadro P1000 both included in a conventional desktop computer.

In order to reach the equilibrium state quickly, we use the method proposed in 
Ref. \citenum{Subramanian2010}. The method starts by locating randomly $C$ monomer 
in a box with the proper dimensions to set the target density. Initially, there 
are $C$ \textit{chains} with one monomer each one. Then, the method adds monomer 
systematically along the chain backbone, rescaling the box properly to conserve 
the density. This process is repeated until the desired chain-length is reached. 
We then run a simulation in which the mean square displacement (MSD) of the 
central monomer is followed and ensure that the chains diffuse enough distance 
to sample the film thickness adequately and the equilibration time is already 
passed before starting to compute observables.
\begin{figure}[h] \centering 
\includegraphics[width=16 cm]{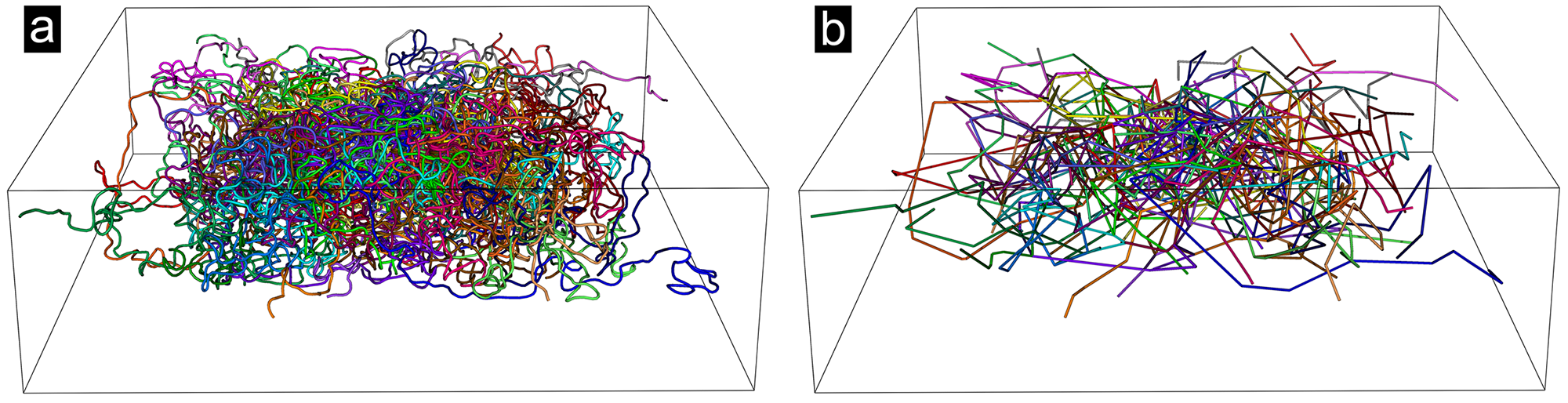} 
\caption{(a) A free-standing film of thickness $h_{eff} \sim 
28$ containing $C=64$ chains of length $N=512$ built with this model. The 3D 
box container is not the simulation box and is just shown to improve 
visualization. The entire chains are shown, without taking into account the 
periodic boundary conditions. (b) Primitive Path chain reduction obtained by 
$Z1$ algorithm. The label color is preserved between original chains and the 
corresponding primitive paths.} 
\label{fig:fig1} 
\end{figure}

\begin{figure}[h]
\centering
\includegraphics[width=7.8 cm]{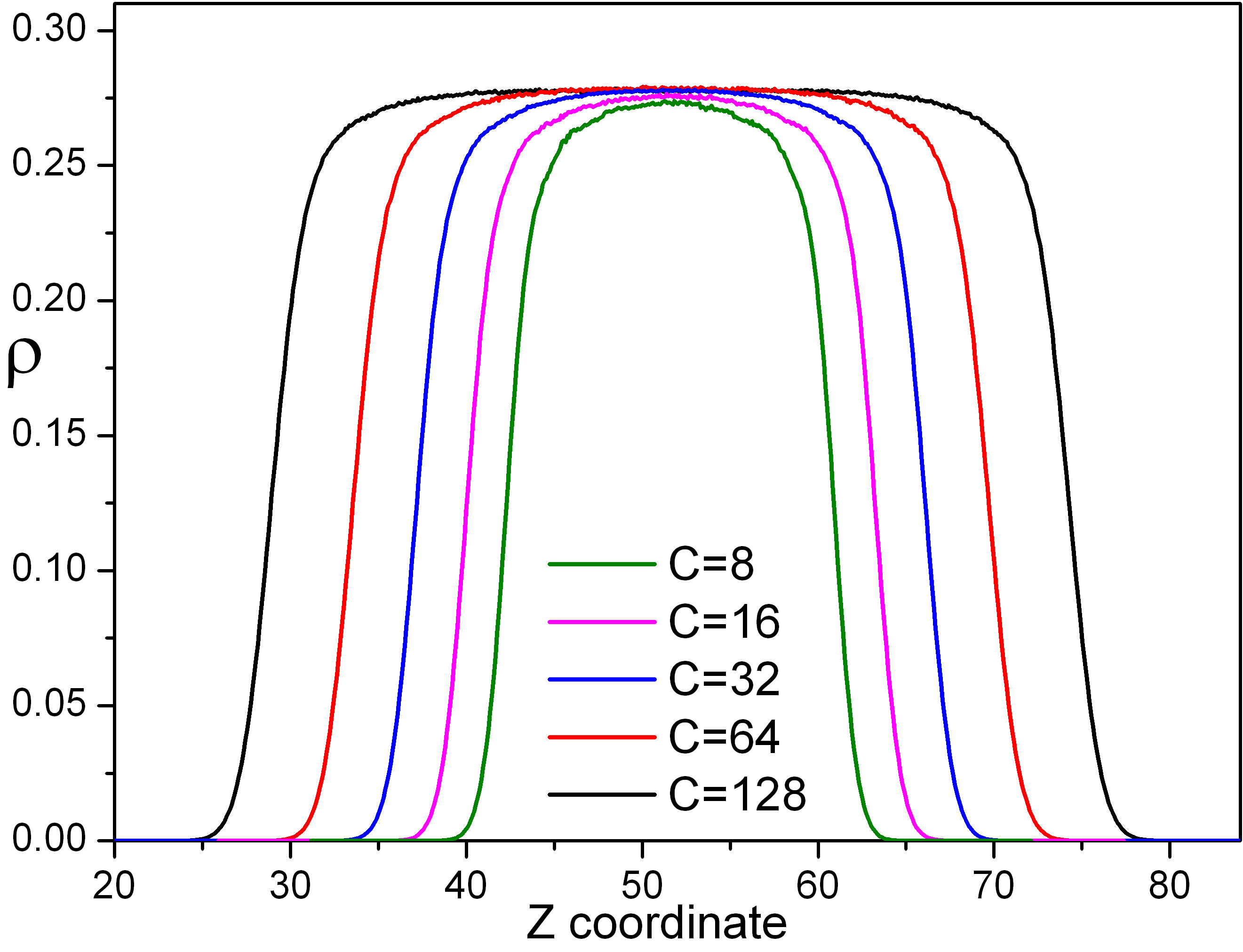}
\caption{Monomer density profiles for a chain length $N=1024$ and films built 
with $C=8, 16, 32, 64$ and $128$ chains.}
\label{fig:fig2}
\end{figure}

The dimensions of the cubic simulation box were chosen to create a system with 
initial monomer density of $\rho_i = 0.12$, so $L_{box}^3 = NC/\rho_i$. As a 
result of the attractive interaction, the system spontaneously forms a thin film 
in a central region of the box, reaching an equilibrium density of 
$\rho_f=0.277$ inside the film. This value is determined from the density 
profile shown in Figure \ref{fig:fig2}. This final density depends on the 
parameter $w$ which, as we mentioned before, is fixed to $w=0.5$ in our study. 
We also confirmed that the largest radius of gyration is roughly three times 
smaller than $L_{box}$, which should be enough to ensure that the chains do not 
interact with their periodic images. Figure \ref{fig:fig1}a shows an 
instantaneous configuration of a free-standing film obtained with this 
preparation protocol.

We have chosen to work with monodisperse linear chains of $N = 512$, $1024$ and 
$2048$ monomers conforming self-confined films containing $C = 8$, $16$, $32$, 
$64$, $128$, $256$, and $512$ chains. Also, in order to have systems of 
reference for the different chains lengths, it was performed simulations of $C=64$ 
chains with $N=512$, $1024$ and $2048$ monomers with the same parameters as 
before but in bulk conditions, i.e., setting the box dimension correctly to get 
a constant density of $\rho_{bulk}=0.277$.

Figure \ref{fig:fig2} shows the monomer density profiles for some films built 
with chains of length $N=1024$. The film thickness is obtained by fitting these 
profiles with the following hyperbolic tangent function as a function of the $z$ 
coordinate:
\begin{eqnarray}
\rho(z) = \frac{\rho_0}{2} \left( 1 - \tanh\left(\frac{\arrowvert z - z_0 \arrowvert - \xi}{d}\right) \right)
\label{Eq:hiperbolic}
\end{eqnarray}
where $\rho_0$ is the density in the interior of the film, $\xi$ is the half 
width of the interior thickness, $z_0$ is the position of the middle film, and 
$d$ is a measure of the width of the interface, which is a consequence of the 
density fluctuations near the surface. Finally, the effective film thickness can 
be reduced to $h_{eff} = 2\xi$. By this measure, the thickness of the profiles 
in Figure \ref{fig:fig2} were $\sim 17.6$, $22.2$, $28.0$, $35.3$, and $44.4$ in 
units of the monomer diameter $\lambda$ and the inner density for all of them is 
$\rho=0.277$. As is expected, these values are in good agreement with the direct 
estimation of the thickness from $h_{eff}=(\rho_i/\rho_f)L_{box}$.

To characterize the behavior of some important vectors (segments of $PP$, 
$\mathbf{R_{ee}}$, etc.), we will use the second Legendre polynomial: 
\begin{eqnarray}
P_2 = \frac{3}{2}\langle\cos^2(\theta)\rangle-\frac{1}{2}
\label{Eq:P2_param}
\end{eqnarray}
where $\theta$ is the angle between the vector under study and a given 
\textit{fixed} direction of interest defined by a unit vector called the 
director, which will be explicitly indicated in each case. $P_2$ is widely used 
to study nematic order in diverse systems (liquid crystal, etc.), and is also 
helpful to quantify the behavior of a vector (or a vector field) around a given 
direction of interest. This order parameter lies within the interval $-0.5 \leq 
P_2 \leq 1$, where a value of $P_2 = 1$ indicates that vectors under analysis 
align perfectly with the direction of reference, $P_2 = 0$ corresponds to an 
isotropic distribution around the reference direction. The negative value of the 
lower bound, $-0.5$, corresponds to vectors all oriented in a plane 
perpendicular to the director.

The topological analysis of the systems presented in the following are all 
performed using the \textit{$Z1$ algorithm} 
\cite{Kroger2005,Shanbhag2007,Karayiannis2009,Hoy2009}, a method based on the MD 
trajectories which finds entanglement by geometrical minimization. In this code, 
all chain ends are maintained them fixed in the space, excluded volume 
interactions are disabled, but the chain uncrossability condition is preserved. 
Then, a set of geometric operations are applied over of all this 
\textit{pseudo}-chains, which monotonically reduce its contour lengths. 
Eventually, the method builds a $PP$ for each chain thereby reducing the linear 
polymer system to an entanglement network of $PPs$. This iterative geometrical 
minimization procedure terminates as soon as the mean length of all $PPs$ has 
converged. Figure \ref{fig:fig1}b shows the $PP$ network obtained by applying $Z1$ 
algorithm to the film plotted in Figure \ref{fig:fig1}a, here is important to 
remark that the chains are drawn entirely, i.e., without cutting at the periodic 
boundary conditions of the simulation box. As a result, some chains (or 
extremes) which appear isolated, have entanglements, as they cross the box 
boundary and are effectively surrounded by periodic images of the chains 
represented.

Additionally, the $Z1$ code provides the statistical properties of the 
underlying topological network but also the positions of the interior 
\textit{kinks}\cite{Shanbhag2007,Karayiannis2009} along the three-dimensional $PP$ 
for each chain. For long chains, the number of kinks is proportional to the 
number of entanglements and in this context, both terms can be considered as 
\textit{equivalents}. In this approach, self-entanglements (intramolecular 
knots) are neglected, as they represent a small fraction and are irrelevant for 
most polymeric systems.

\section{Results and discussion}

\subsection{Statistics of entanglements in bulk}
\hspace{\parindent}
As a reference, we start by using the $Z1$ 
algorithm\cite{Kroger2005,Shanbhag2007,Hoy2009,Karayiannis2009} to perform a 
topological analysis of the bulk system configurations. 

\begin{figure}[h]
\centering
\includegraphics[width=8.6 cm]{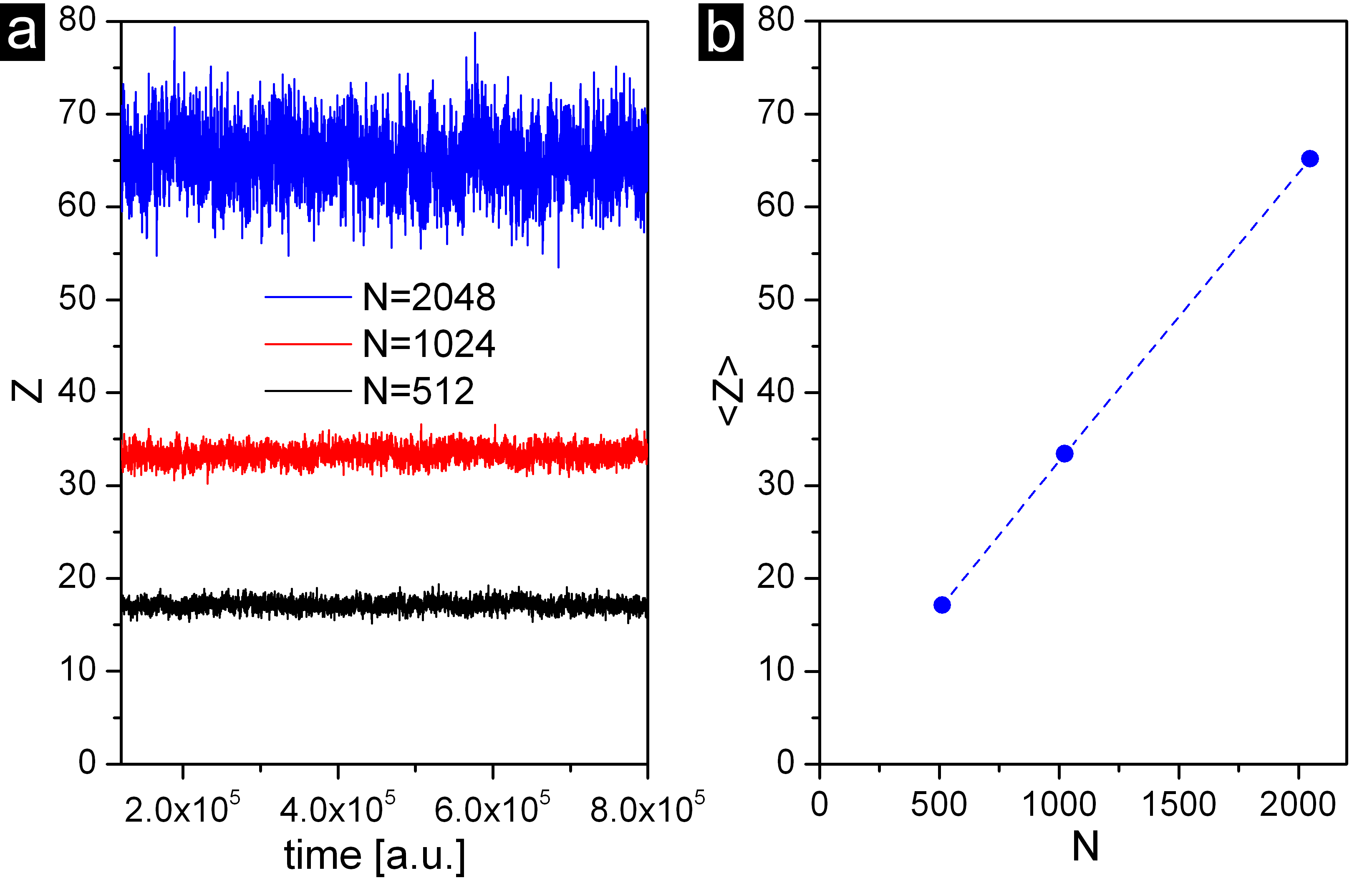}
\caption{
(a) Temporal evolution of the average number of entanglements per chain ($Z$) 
measured by the $Z1$ algorithm after equilibration {for three different melts 
built with $C=64$ chains of lengths $N=512$, $1024$ and $2048$} in bulk 
conditions at $\rho_{bulk}=0.277$. (b) Temporal average number of entanglements 
per chain $\langle Z \rangle$ as a function of the chain-length $N$.
}
\label{fig:fig3}
\end{figure}

Figure \ref{fig:fig3}a shows the temporal evolution of the total numbers of 
entanglements per chain ($Z$) after equilibration. This number fluctuates slightly 
around its average value, which is an indicator that the systems were well 
equilibrated. Moreover, as is expected in bulk at fixed density, the average 
value $\langle Z \rangle$ of the number of entanglements per chain increases 
linearly with the chain-length $N$, with a slope of $\alpha = 0.03125$ (Figure 
\ref{fig:fig3}b). This slope is just the reciprocal of the entanglement length 
$N_e = 1/\alpha = 32$ (in number of monomers), which can be used to characterize 
the crossover between the Rouse and Reptation regimes. In the reptation model, 
$N_e$ is defined as the arc length of a chain with mean-square end-to-end 
distance equal to tube diameter $a$ ($N_e=(a/b)^2$, $b$ being the statistical 
segment length).

The statistic of entanglements in bulk systems is well explained by the 
\textit{chain packing model}\cite{Lin1987,Kavassalis1988,Fetters1994}. Mostly, 
the idea is that the larger the dimensions of a chain, the greater the volume 
pervaded by that chain, so the greater the number of other neighbors chains it 
will encounter and with which it might entangle. In this model, $N_e$ is defined 
as the ratio of the pervaded volume $V_p$ to the real volume occupied by the 
chain $V_c$. Although $V_p$ is not easy to calculate, a well-accepted estimate 
is proportional to the volume covered by one of their characteristic lengths of 
the chain: $R_{ee}$ or $R_g$. Thus, the pervaded volume can be estimated as $V_p 
\propto R_{ee}^3$ while the effective volume occupied by the chain is $V_c 
\propto N\lambda^3$. This model assumes that an entanglement arises when the 
molecular weight and the concentration are such that at least two chains share 
the same pervaded volume, i.e., $V_p/V_c \sim 2$.

These volumes scale differently with molecular weight, $V_p \propto N^{3/2}$ 
while $V_c \propto N$. As a result, increasing the chain length increases the 
number of chains that are allowed to share the same pervaded volume, favoring 
the interchain contacts that lead to entanglements. In the following, we will, 
however, see that this model is not sufficient to explain the statistics of 
entanglements in confined systems.

\subsection{Global impact of the confinement on the entanglements}
\begin{figure}[h]
\centering
\includegraphics[width=8.0 cm]{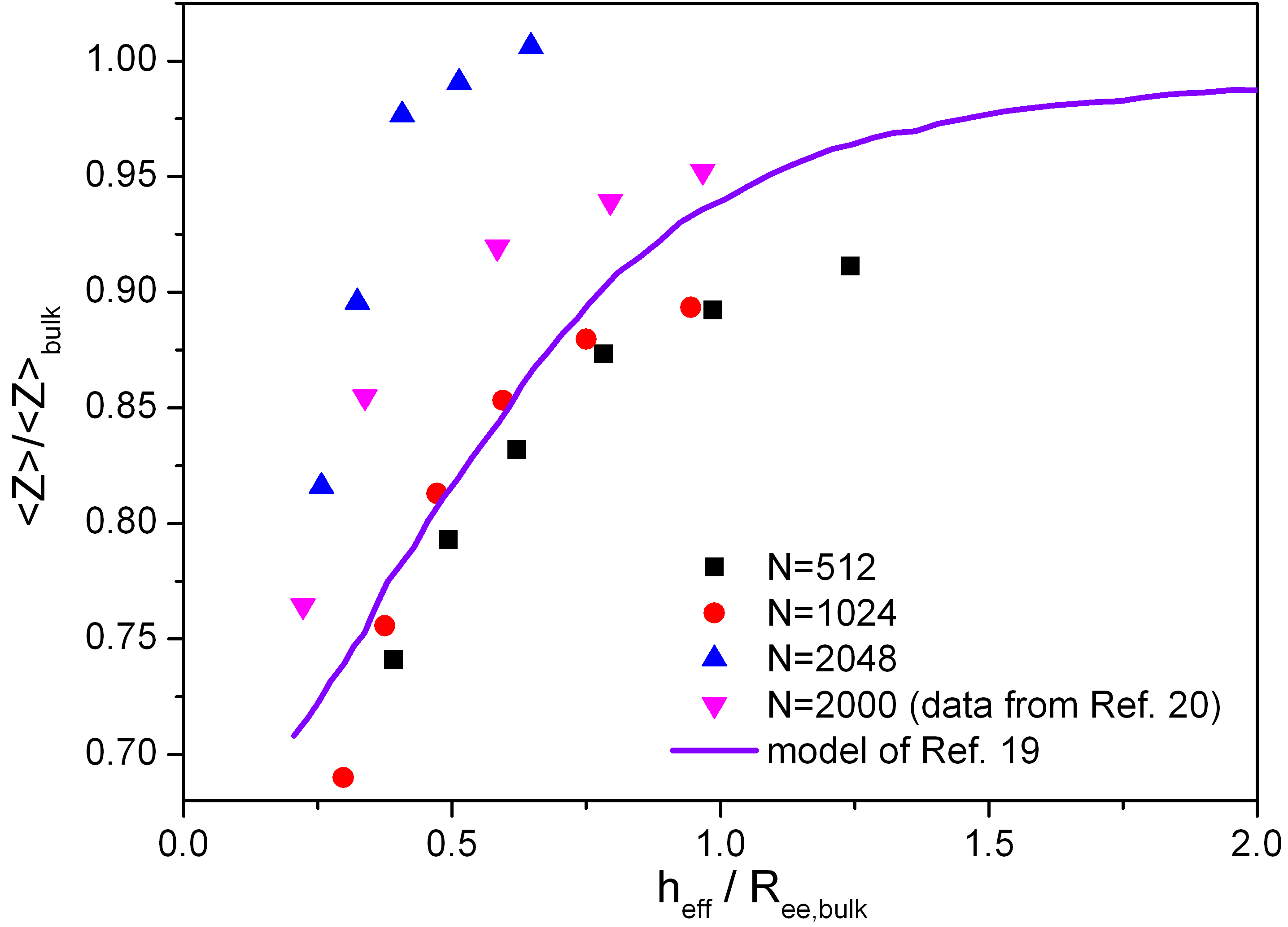}
\caption{
Normalized reduction of entanglements per chain as a function 
of confinement for all free-standing films thicknesses studied here, for 
chain lengths $N=512$, $1024$ and $2048$. Data of a similar system reported in 
Ref. \citenum{Sussman2016} and the model proposed in Ref. \citenum{Sussman2014} are 
also included.
}
\label{fig:fig4}
\end{figure}

\hspace{\parindent}
Figure \ref{fig:fig4} shows the total number of entanglements per chain compared 
to its bulk value, $\langle Z \rangle / \langle Z \rangle_{bulk}$, for thin 
films of various thickness. The film thickness $h_{eff}$ is normalized here 
using the average end-to-end distance in bulk conditions for the same chain 
length. This normalization is guided by the proposal of reference 
\cite{Sussman2014}, which suggests that the corresponding curve should be 
universal in the limit of large molecular weight.

Globally, we find that confinement leads to a decrease in the average number of 
entanglements per chain. Qualitatively, this result is in good agreement with 
the one observed in experiments\cite{SiLun2005,Rathfon2011,LiuYujie2015}, 
simulations\cite{Cavallo2005,Vladkov2007,Ramirez2015}, and with the theoretical 
model proposed in Ref. \citenum{Sussman2014}. However, Figure \ref{fig:fig4} shows 
that, quantitatively, there is a notable difference between our results and the 
model proposed by Sussman and coworkers\cite{Sussman2014}. Figure \ref{fig:fig4} 
clearly shows that the model accounts reasonably well for the data obtained for 
shorter chains, but strong deviations are observable for thin films made of long 
chains. In those films, the decrease of the entanglements is observed only for 
films that are significantly thinner than the size of the unperturbed chain. 
Data for $N=2000$ extracted from a recent manuscript\cite{Sussman2016} displays 
a similar trend. In the next section, we discuss possible reasons for the origin 
of the discrepancy between theory and simulation.

\begin{figure}[h]
\centering
\includegraphics[width=7.8 cm]{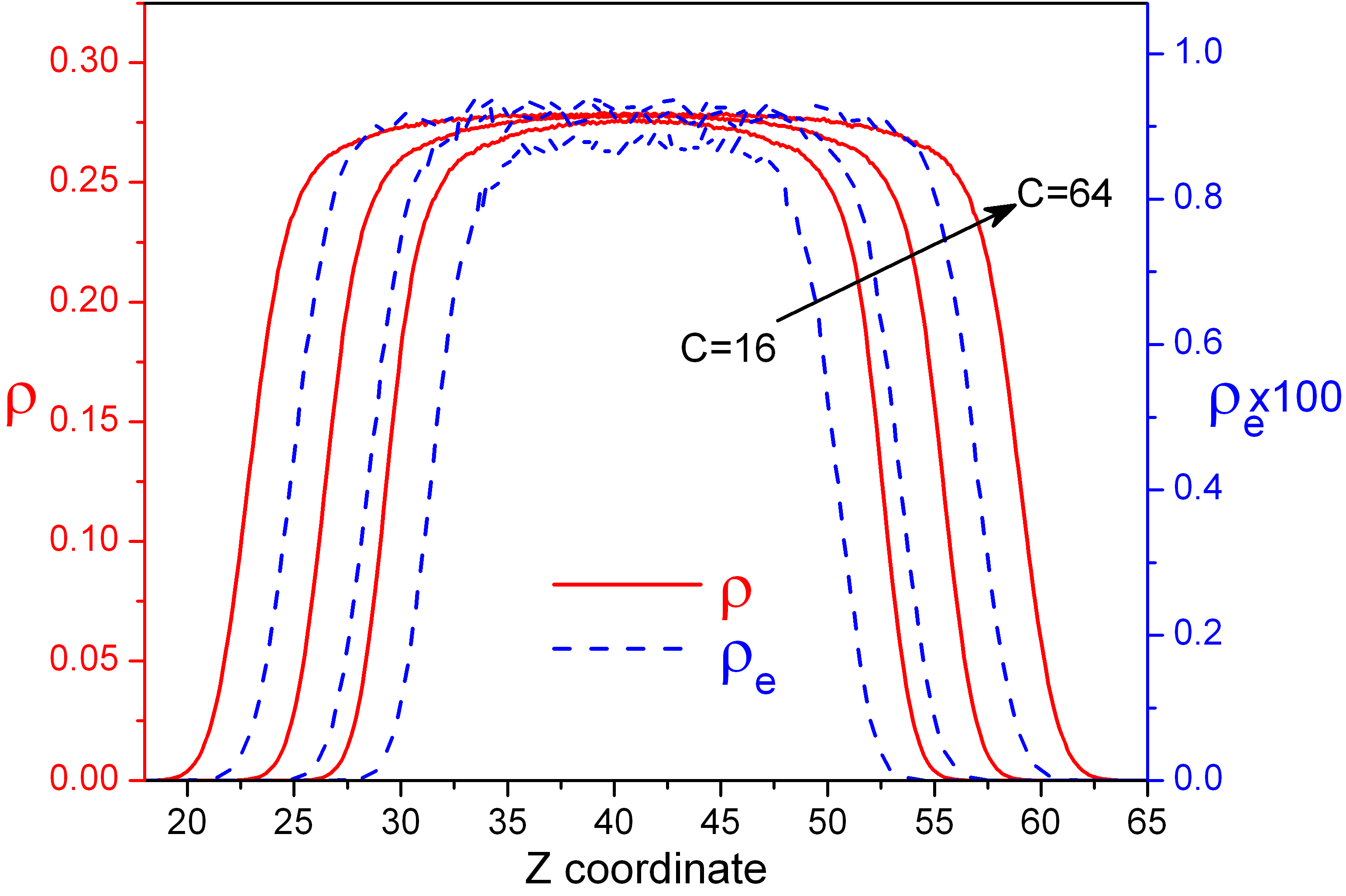}
\caption{
Monomer and entanglement density profiles. For $N=1024$ and three films of 
$C=16$, $32$ and $64$ chains respectively. Two different scales (left blue and 
right red axis) are used to plot these quantities in the same figure for 
comparison.
}
\label{fig:fig5}
\end{figure}

The $Z1$ algorithm also provides the primitive path ($PP$) conformation of each 
chain, from which it was possible to determinate the position of entanglements 
within the film. A spatially resolved profile of the entanglement density across 
the film is shown in Figure \ref{fig:fig5} and compared with the monomer density 
profile for films of different thickness built with chains of $N=1024$ monomers.

Is interesting to note in Figure \ref{fig:fig5}, that entanglements sample the 
space uniformly within the film exhibiting a notable decrease only near to the 
surface.

\subsection{Understanding the discrepancy between simulations and theory}
\hspace{\parindent}
The theoretical model presented in Ref. \citenum{Sussman2014} is based on three 
fundamental hypothesis: (I) validity the principle of conformational transfer 
proposed by Silberberg\cite{Silberberg1982}, generalized to a thin film 
geometry, (II) the distribution of orientations of the end-to-end vector is made 
anisotropic by the geometric confinement, and this orientation distribution is 
directly communicated to the primitive path network, (III) the distribution of 
orientations at the $PP$ scale is used to predict the changes in the 
entanglement network. In the following, we will analyze the validity of these 
assumptions for our simulations, in order to understand the origin of the 
observed discrepancy. 

Hypothesis (I), involves a modification of the Silberberg 
model\cite{Silberberg1982} which treats the chain as a random walk using 
reflecting boundary conditions to compute changes to the chain conformation in 
the presence of a wall. The original model formulated by Silberberg consider the 
perturbation of chains near to one wall in space and makes quantitative 
predictions for the statistics of the chain conformations in the direction 
normal to the surface.

Following the philosophy of Silberberg, the authors of Ref. \citenum{Sussman2014} 
proposed an extension of this idea for thin films. Two walls with reflecting 
boundary condition delimit the film, and the contribution of both surfaces are 
added to obtain the chain conformation inside the film. At first order, the 
early two reflections are taken into account. Formally, in analogy with the 
method of images in electrostatics, second and higher order reflections should 
also be taken into account, so that the final results involves summing an 
infinite series. Fortunately, due to the fast decaying of the superior order 
contributions, the convergence of the series is quick, and only a few terms are 
needed to reach an accuracy. This extended model allows one to predict the 
change in the normal component of the mean end-to-end vector as a function of 
the distance from the final random walk step to the surface, $R_{ee,z}^2(z)$. 
Integrating this function through the film thickness, it is possible to compute 
the global change in the normal component as a function of the film thickness. 
For further information on this approach see Ref. \citenum{Sussman2014}.

\begin{figure}[h]
\centering
\includegraphics[width=8.0 cm]{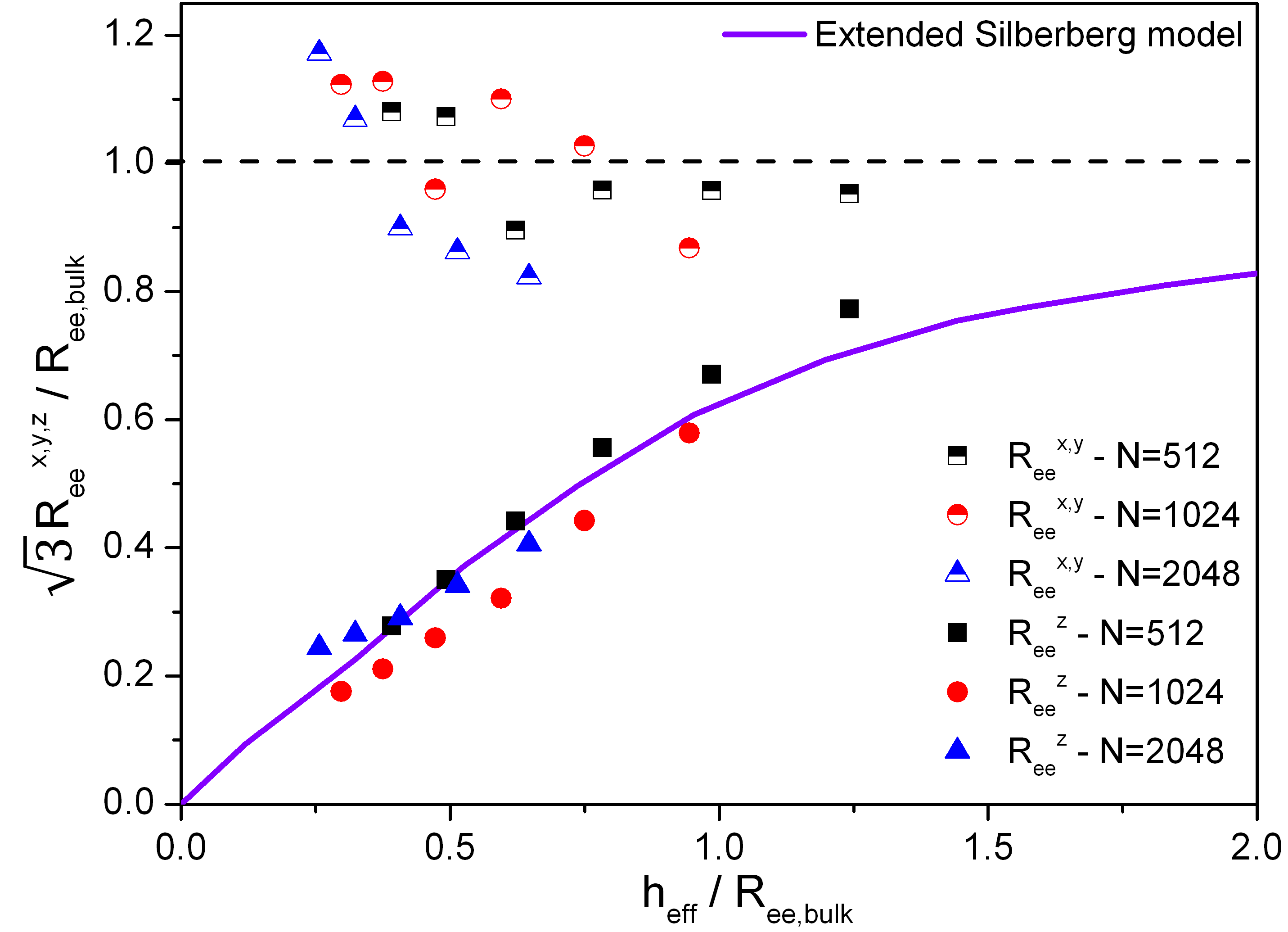}
\caption{Root-mean-square components of the end-to-end as a function of the 
normalized film thickness for all thin films and chain lengths studied here. 
Filled symbols are the component normal to the surface of confinement $R_{ee}^z$ 
for different chain-lengths, and the semi-filled symbols are the components 
parallel to the surface, $R_{ee}^{x,y}$. The continuous line is the prediction 
of the extended Silberberg model {and the dashed black line is the bulk value 
$R_{ee,bulk}/\sqrt{3}$.}
}
\label{fig:fig6}
\end{figure}

In Figure \ref{fig:fig6}, the prediction of this extended Silberberg model for the 
perpendicular component of the bulk normalized mean end-to-end vector as a 
function of the normalized films thicknesses is compared with our simulation 
results for different chain length exhibiting an excellent agreement and 
supporting the validity of this first hypothesis. 

Figure \ref{fig:fig6} also displays the normalized parallel components ($R_{ee}^x$, 
$R_{ee}^y$) averaged over the film thickness. As expected, these components are 
only {slightly} affected by confinement and exhibit a bulk-like behavior.

\begin{figure}[h]
\centering
\includegraphics[width=8.6 cm]{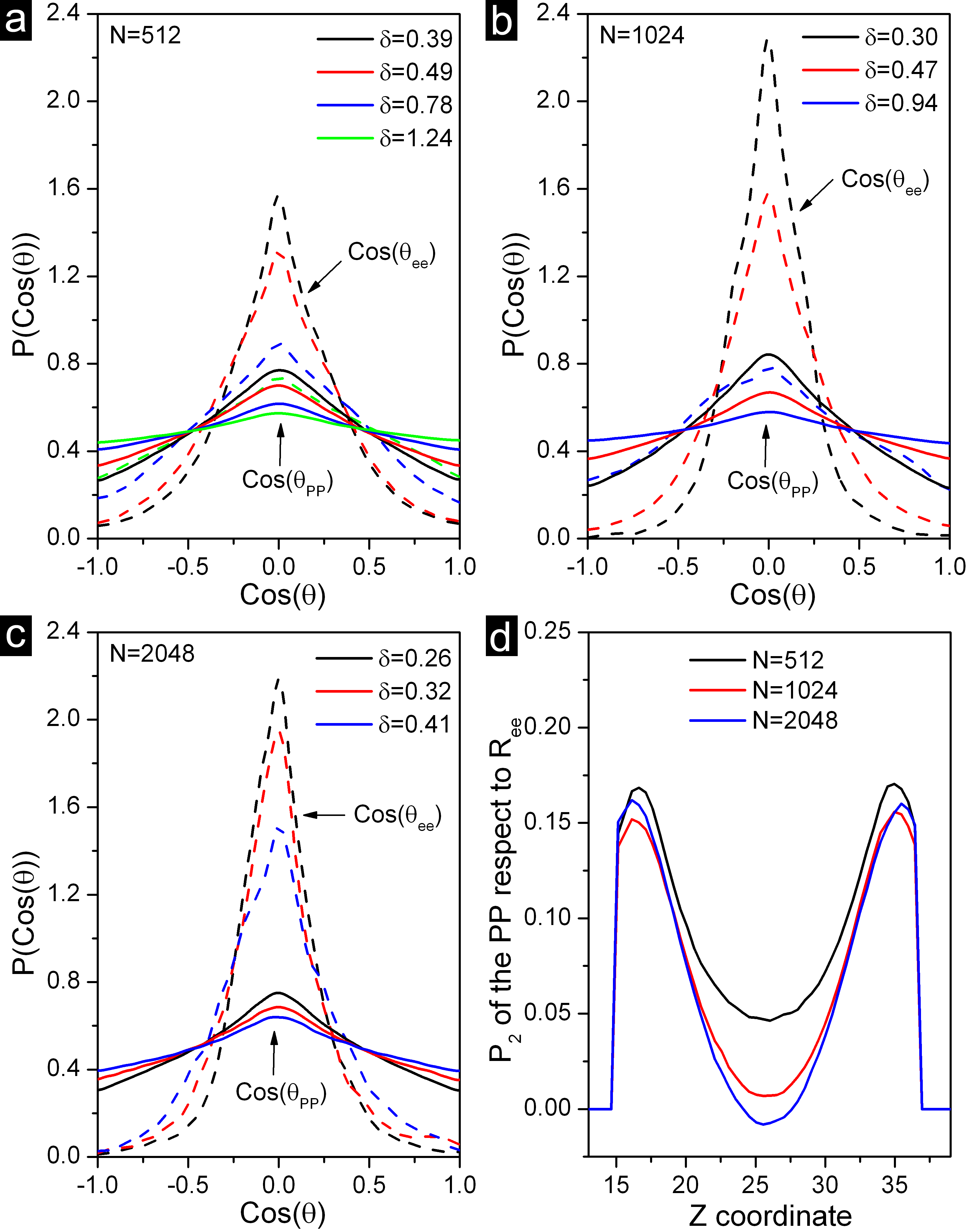}
\caption{
(a), (b), (C) Average distribution of the angle with respect to the $z$-axis of 
the $PP$ segments (solid lines) and the $R_{ee}$ vectors (dashed lines) {for 
different chain lengths and confinement strength $\delta$.
} 
(d) profile of the $P_2$ parameter evaluated using the angle defined between the 
$PP$ segments and the $R_{ee}$ vector.
}
\label{fig:fig7}
\end{figure}

In order to check the second hypothesis (II), we analyzed the orientational 
probability distributions of the $PP$ segments and $R_{ee}$ vectors concerning 
the $z$-direction for different degrees of confinement, here expressed as 
$\delta = h_{eff}/R_{ee}$. These results are shown in Figure \ref{fig:fig7}a, b 
and c where is immediately evident, at least globally speaking, the orientations 
present $R_{ee}$ are not communicated to the $PP$ length scale, shedding doubt 
on the validity of this second hypothesis. 
The chains are much more strongly oriented at the scale of the end-to-end vector 
than at the scale of the primitive path segments. The data in Figure 
\ref{fig:fig7}a shows that as the confinement increases ($\delta$ decreases) 
the two distribution become peaked around zero, i.e., the chain tend to lie 
parallel to the interface. However, the order of magnitude of the effect is much 
more pronounced for the end-to-end vector than for the primitive path segments.

Furthermore, this difference becomes more notable for longer chains. The 
distributions for $R_{ee}$ becomes more peaked for $N=1024$ 
(Figure \ref{fig:fig7}b) and even more for $N=2048$ (Figure \ref{fig:fig7}c) for a 
similar degree of confinement, while the orientation of the primitive path seems 
insensitive to the chain length and only slightly dependent on $\delta$.

\begin{figure}[h]
\centering
\includegraphics[width=15.0 cm]{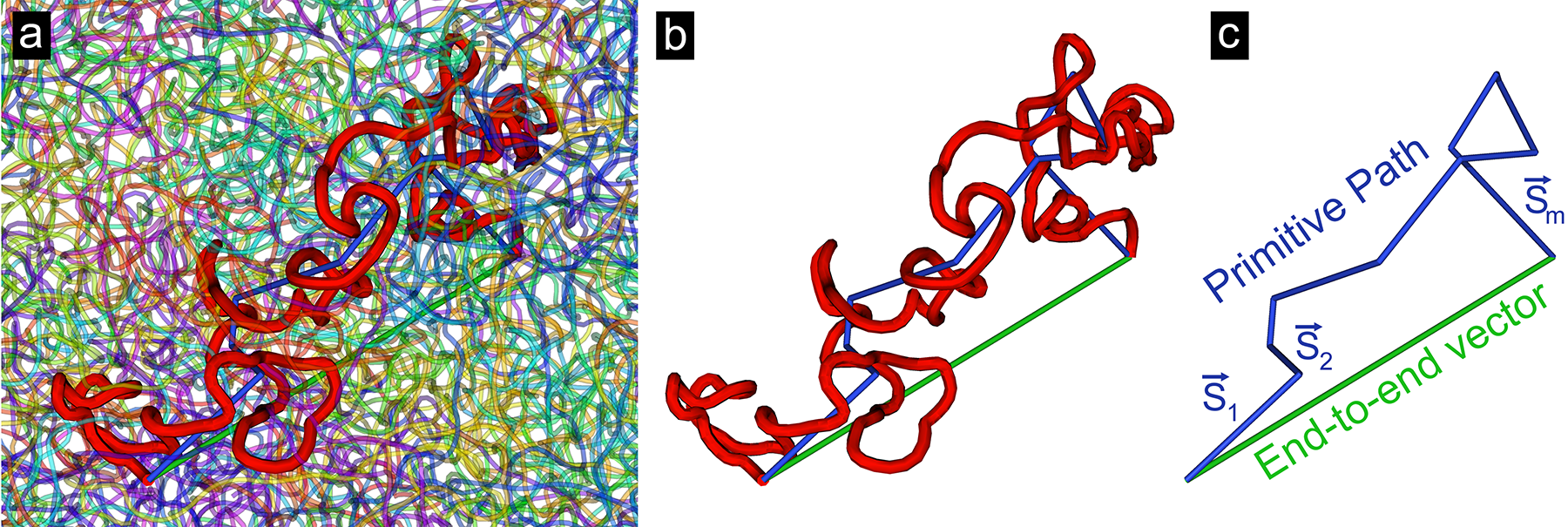}
\caption{
The method implemented to study the local orientational correlation between the 
$PP$ segments and the end-to-end vector through $P_2$. In all those images are 
shown the PP (in blue) and the end-to-end vector (in green). (a) Highlighted in 
red, a chain embedded in the film. All the others chains are thinned and 
transparent. In (b) all others chains were removed to improve visualization. (c) 
The segments $S_1,S_2,..., S_m$ forming the $PP$ are used to evaluate $P_2$ 
relative to the end-to-end vector.
}
\label{fig:fig8}
\end{figure}

In addition to this global analysis, we have studied the local behavior of both 
vectors to assess the possible existence of a local correlation. We have 
evaluated a profile of the $P_2$ order parameter for the angle between the 
end-to-end vector of a chain and the $PP$ vectors (segments) belonging to this 
chain. {The sketch in Figure \ref{fig:fig8} illustrates this idea 
and Figure \ref{fig:fig7}d reports this observable.}

The local correlation becomes more important near the surface, however, all 
values remain below $P_2=0.3$, indicating a very poor orientational correlation 
between these vectors. Moreover, the range over which the correlation is felt 
appears to be independent of chain length, and for longer chains, the 
correlation is completely lost in the middle of the film.

Clearly, these observations indicate that, while the extension of the Silberberg 
model gives an accurate picture of the global chain conformation, the primitive 
path is much less affected by confinement than expected in the theory. Indeed, 
the thickness over which the orientation of the primitive path is affected does 
not appear to scale with molecular weight but is restricted to a finite 
thickness layer at the surface of the film. This leads to the deviation of 
$\langle Z \rangle/\langle Z \rangle_{\mathrm{bulk}}$ from the 
$h_{\mathrm{eff}}/R_{ee}$ scaling reported in figure \ref{fig:fig4}.

\subsection{Chains conformations and the importance of surface effect}
\begin{figure}[h]
\centering
\includegraphics[width=8.6 cm]{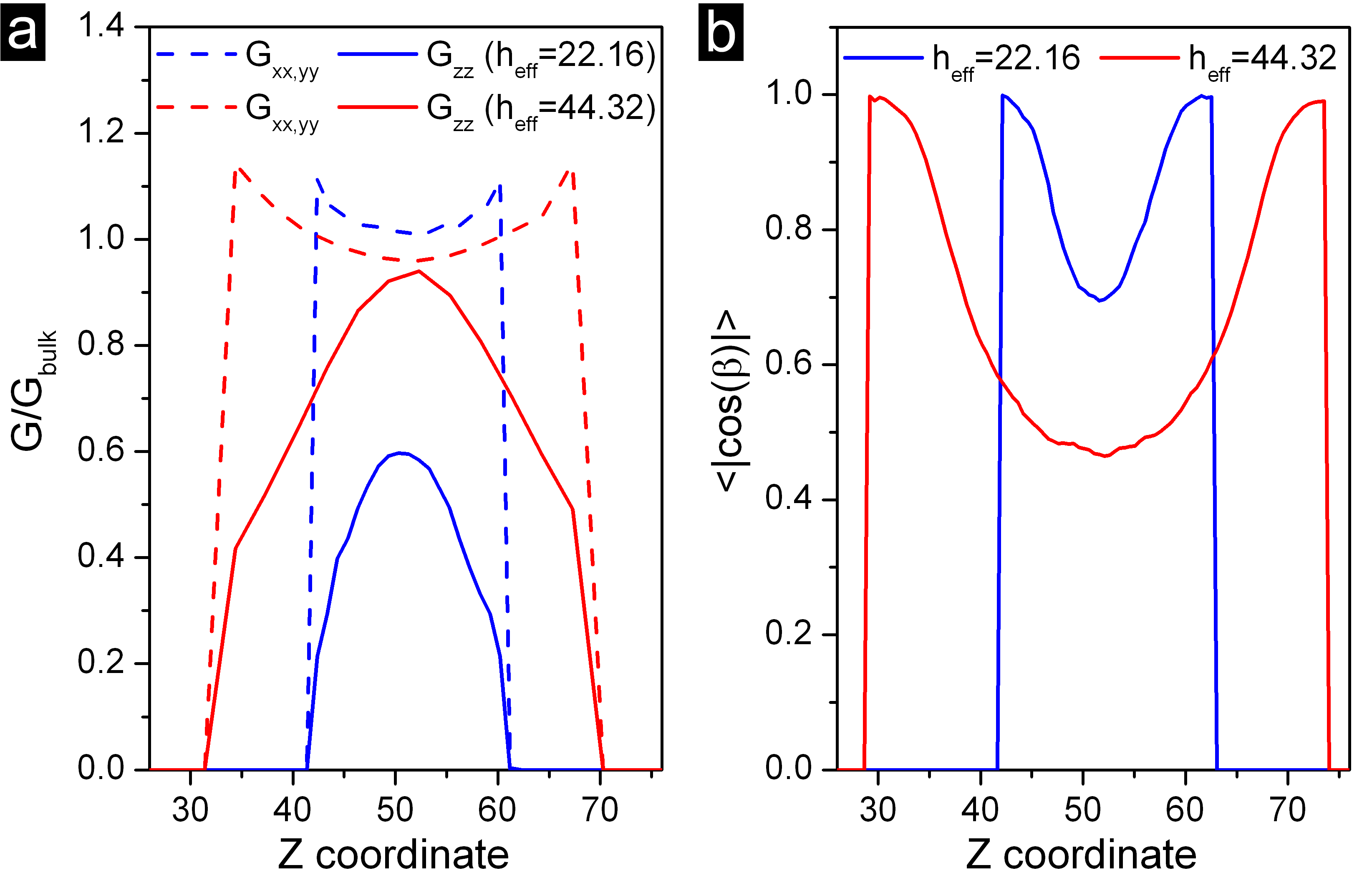}
\caption{
Profiles evaluated using the location of the center of mass of the chains within 
the film of (a) Bulk normalized and temporal-averaged components of the radii of 
gyration $G_{xx,yy}$ and $G_{zz}$, (b) Average-absolute value of the cosine 
director of the eigenvector associated with the minimum eigenvalue of the 
gyration tensor with respect to the surface of confinement. 
{These observables were evaluated for two film thicknesses 
$h_{eff}=22.16$ (in blue) and $h_{eff}=44.32$ (in red).}
}
\label{fig:fig9}
\end{figure}

\hspace{\parindent}
Heretofore, we have studied how the confinement affects the statistics of 
entanglements and the distributions of end-to-end vectors or primitive paths 
segment self-confined free-standing films. In this section, we investigate how 
the confinement alters the global shape of the chains.

The first set of descriptors are the three diagonal components of the inertia 
tensor (or gyration tensor) $G_{xx}$, $G_{yy}$ and $G_{zz}$. The squared 
gyration radius can be expressed as $R^2_g=G_{xx}+G_{yy}+G_{zz}$.
Figure \ref{fig:fig9} shows the averaged value of these components normalized with 
the corresponding bulk value as a function of the position of the center of mass 
of the chain. To avoid redundancy of data, we report these observables for only 
two different thicknesses ($h_{eff}=22.16$ blue lines, $h_{eff}=44.32$ red lines) 
built with chains of $N=512$ monomers, but the data obtained with varying 
lengths of chain are very similar. 

In Figure \ref{fig:fig9}a is possible to see that for both thicknesses (continuous 
lines) the component $G_{zz}$, associated with the direction of confinement, 
tends to induce a noticeable shrinking of the chains on that direction, i.e., 
the chains break their spatial isotropy and tend to adopt a flat shape near the 
surface. In counterpart, the components $G_{xx,yy}$ slightly change their 
values, increasing as much a $10\%$ compared to their bulk value (dashed lines 
in Figure \ref{fig:fig9}a), which is in good agreement with previous 
works\cite{Muller2002}. 

Until now, we have used the tensor as is, expressed in the canonical system of 
reference $\{x,y,z\}$, i.e., without diagonalizing in its principal axes. By 
diagonalizing the $G$ tensor and studying the cosine director of the eigenvector 
associated with the minimum eigenvalue (which represent the most important 
direction to where the chain is elongated), it is possible to know the main 
direction of this flatness. In Figure \ref{fig:fig9}b is reported how the 
average orientation of this vector is dictated by the position of the chain 
within the film. Independently of the molecular weight, all chains are flat at 
the edges of the film, then this effect decreases monotonically while entering 
in the film. This idea is also in good agreement with the predicted compression 
of the end-to-end vector discussed in the previous section.

In Figure \ref{fig:fig9}a and Figure \ref{fig:fig9}b it is seen that the 
flattening alters the chain shape all across the film thickness for this 
chain-length ($N=512$, $R_g \sim 19.1$). In Figure \ref{fig:fig9}a, at the edge 
of the thicker film ($h_{eff}=44.32$, red lines) the chain compression reaches a 
maximum of around $60\%$ compared to its bulk value as is evidenced in the 
component of the radius gyration perpendicular to the plane of confinement 
($G_{zz}$). Then, the compression decreases monotonically along the film 
achieving a bulk-like state in the center. Considering here that the film 
thickness is around two times $R_g$ (i.e., the chain size is comparable with the 
film thickness) it is reasonable that the chain whose center of mass is located 
roughly in the middle film adopt bulk-like conformations.

In the case of the thinner film ($h_{eff}=22.16$, blue lines) the chains 
experience a stronger shrinkage at the boundary, and although the effect 
decreases inside the film, the bulk state is not reached in the center. Which is 
also reasonable due to this thickness is almost the half of the chain size 
($2R_g$), so the chain conformation is strongly confined. Even in the center of 
the film the chain reaches only around half of its bulk size in the 
$z$-direction. This simple analysis provide an accurate picture of the chain 
conformations across the film.

\subsection{Primitive path network under confinement}
\begin{figure*}[h]
\centering
\includegraphics[width=17 cm]{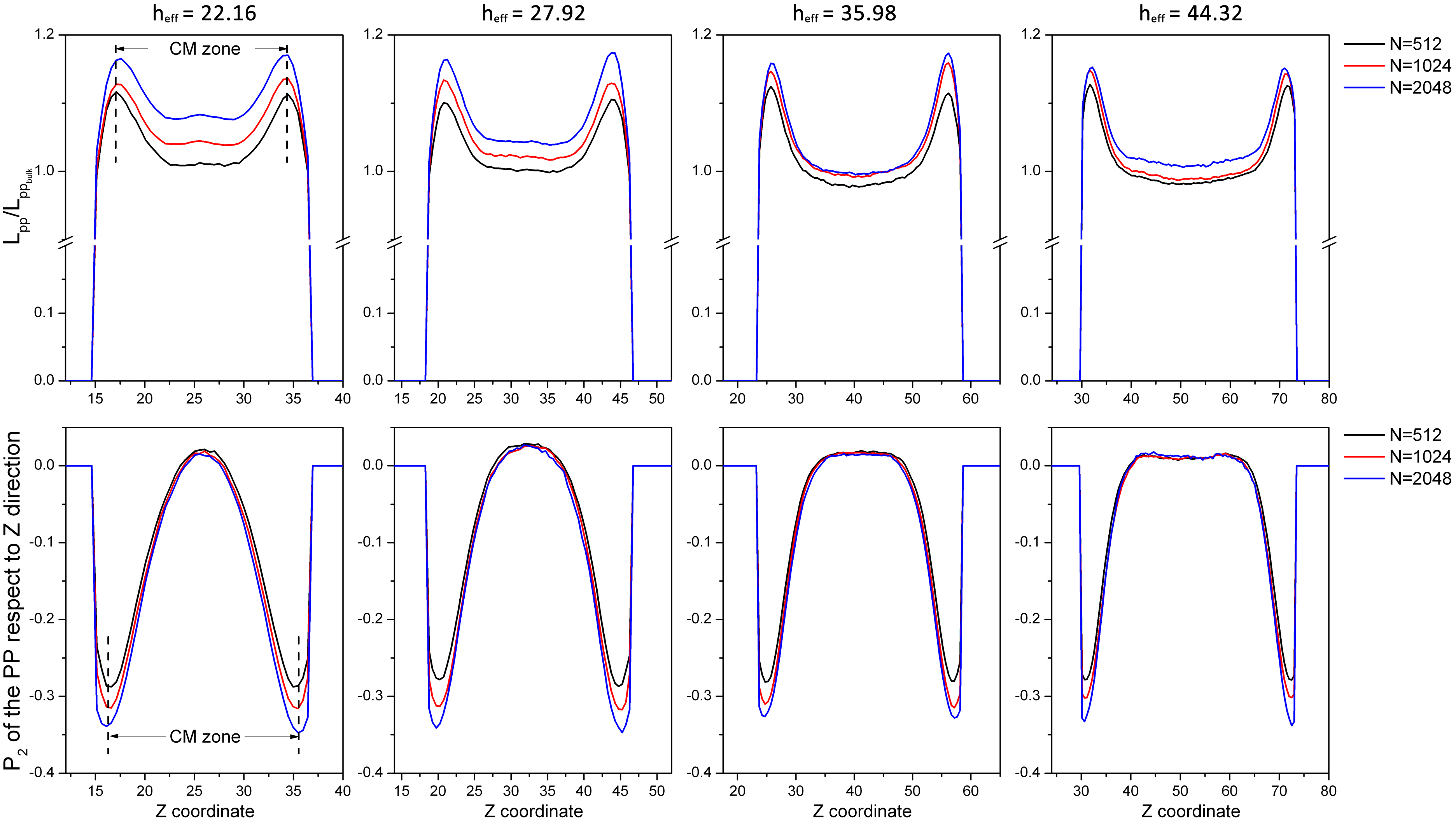}
\caption{Primitive path characterization. In the top, average $PP$ segment 
length ($L_{pp}$) normalized with the bulk value. Bottom, $P_2$ order parameter 
evaluated on the angle defined between the $PP$ segments and the 
$\hat{z}=(0,0,1)$ direction. The position in the profile is computed using the 
geometrical center of the $PP$ segments. The zone delimited by the center of 
mass of the chains is indicated.}
\label{fig:fig10}
\end{figure*}

\hspace{\parindent}
In this section, we study how the confinement impacts the primitive path 
network. To quantify this confinement effect, we have calculated the profiles 
across the film of two characteristic quantities of the $PP$ segments. One is 
the profile of the $P_2$ order parameter for the angle between the $PP$ segments 
and the normal to the plane of confinement ($\hat{z}=(0,0,1)$) and the other one 
is the length of the $PP$ segments. The location within the film was computed 
using the geometrical center of the $PP$ segments.

Here, is relevant to mention that we just will consider these observables inside 
of the called \textit{``center of mass zone''} (marked as CM zone in Figure 
\ref{fig:fig10})), i.e., the zone within the film reachable by the centers of 
mass of chains. Due to the nature of the $PP$ segments, they can exist beyond 
the space reachable by the center of mass of the chains and that is the reason 
of why $L_{pp}$($P_2$ of $PP_z$) after achieving a maximum(minimum) goes to 
zero. We note that the location of these extrema coincides with the boundary of 
the center of the mass zone. Beyond this limit, the data is mostly related to 
the tails of the chains.

Interestingly, we found that the primitive path segments are quite insensitive 
to the confinement and the chain-length exhibit two characteristic \textit{weak} 
response depending on their location within the film. There is a relatively vast 
region of the center-film where the length of the segments is constant, and at 
some point, near to the edge, their length increases monotonically, stretching 
up to $10\%-18\%$. We note that the starting point for these deviations is at a 
distance of around one segment length of $L_{pp,bulk}$ from the edge. This seems 
reasonable if we consider that the segments contributing statistically to this 
part of the data are in the zone of maximum chain compression and, as we 
explained before, decreases the number of entanglement locally increasing at the 
same time $PP$ segments length. Furthermore, this compression forces the $PP$ to 
align parallel to the surface of confinement as seen in the bottom graphs 
showing $P2$ in figure \ref{fig:fig10}, where the negative number indicates 
perpendicularity with the $z$-direction.

This behavior seems universal for the $PP$ network and only weakly dependent on 
the confinement strength and the molecular weight. Only for strong confinement 
($h_{eff} = 22.16$, $h_{eff} = 27.92$), the curves obtained for different 
molecular weights slightly depart from each other by a small vertical shift.

\section{Summary and conclusions}
\hspace{\parindent}
In summary, we have performed an analysis of entanglement statistics in a 
coarse-grained model of free-standing thin films made out of long linear 
polymers.

We found that the geometric confinement breaks the isotropic conformation of the 
chains, compressing and flattening their shape in the direction perpendicular to 
the plane of confinement as is shown in Figure \ref{fig:fig9}a and 
\ref{fig:fig9}b. This anisotropic contraction seems not completely compensated 
in the other direction, and their lateral extension increases by just around 
$10\%$ ($G_{xx,yy}$ in Figure \ref{fig:fig9}a), which results in an effective 
decrease of the volume pervaded by the chain. 

This decrease in the pervaded volume reduces the number of neighbor chains 
inside the shared volume, lowering the potential contacts between them, with as
chief consequence the effective reduction of entanglements, while the monomer 
density remains constant.
However, the flattening effect is poorly captured by the $PP$ network, as is 
reported in Figure \ref{fig:fig10}. First, for all chain length under 
\textit{weak} confinement ($h_{eff}=44.32$), the $PP$ segments seems to be 
unaffected and behave in a bulk-like manner for a wide range in the center of 
the film. It is only near to the surface that the flattening becomes more 
important, inducing the segments to becomes parallel to the surface of 
confinement and increasing slightly (around a $15\%$) their length. This 
\text{pronounced} change in both observables takes place when the $PP$ is within 
a distance comparable to $L_{pp,bulk}$ from the surface. 

In the first two graphs of $L_{pp}/L_{pp,bulk}$ in Figure \ref{fig:fig10} it is 
notable how, under strong confinement, ($h_{eff}=22.16-27.92$) only the $PP$ 
segments associated with longer chains shift their length slightly (just around 
$~10\%$ with respect to the bulk value) inside the film while conserving the 
characteristic effect of increasing near to the film edge.

We also performed a comparison of our data with the theory proposed in Ref. 
\citenum{Sussman2014}, which models the entanglement reduction in confined 
systems as a function of the strength of confinement. After detailed tests of 
the central hypothesis, we found evidence that the second hypothesis: 
\textit{``the oriental correlation at the end-to-end vector scales created by 
geometric confinement are directly communicated to the primitive path 
network''}, seems not to be right. In fact, we found substantial evidence that 
these vectors have uncorrelated orientations, except for a thin layer close to 
the surface, and this effect is even more notable for longer chains. However, 
the extension of the Silberberg model proposed by the same authors fits the 
results for the chain conformation quite well in all the simulated range. A 
better understanding of the response of the primitive path to global chain 
deformation would be desirable to generalize the ideas of Ref. 
\citenum{Sussman2014}.

\section{Appendix I}
As mentioned in section II, since the interparticle potential used in this study 
is built with a soft-core repulsion and an attractive tail, depending on the 
relative weight of both interaction a problem of thermodynamic stability may 
arise. Thus, to ensure the stability of our system it was necessary to determine 
a safety range of values for the \textit{independent parameter} $w$ (see 
Eq. \ref{Eq:SoftPotential}).

Originally, the theoretical framework for predicting the stability of these 
kinds of systems was provided by Fisher and 
Ruelle\cite{FisherRuelle,Ruelle1999}. According to Proposition 3.2.2 in Ref. 
\citenum{Ruelle1999} the stability is ensured if the total potential energy $U$ of 
a given system with $N_t$ particles interacting through a pair potential 
$\Phi(r)$ satisfies the inequality:
\begin{eqnarray}
U(\mathbf{r}_1,\mathbf{r}_2,...,\mathbf{r}_{N_t}) = \sum^{{N_t}-1}_{i=1}\sum^{N_t}_{j>i} \Phi(\rvert \mathbf{r}_i - \mathbf{r}_j \rvert) \geq -N_t\varepsilon
\label{eq:EnergyConvergency}
\end{eqnarray}
where $\mathbf{r}_i$ is the vector position of the particle $i$, and 
$\varepsilon \geq 0$ is a finite constant independent of $N_t$. This inequality 
ensures the convergence of the grand partition function.

Beyond this formal criterion, Fisher and Ruelle also provided two more 
straightforward rules which help one to decide whether a potential will lead to 
a steady thermodynamic state.
The first criterion is a weaker condition:
\begin{eqnarray}
\int d\mathbf{r} \, \Phi(r)>0
\end{eqnarray}
which is necessary but not sufficient, i.e., if $\int d\mathbf{r} \, \Phi(r)<0$ the 
system is unstable.

A sufficient condition for stability is that given in Ref. \citenum{FisherRuelle}
\begin{eqnarray}
\widetilde{\Phi}(k) = \frac{1}{(2\pi)^3}\int d\mathbf{r} \, \Phi(r) e^{-i\mathbf{k} \mathbf{r}} \geq 0
\end{eqnarray}
with the following equivalent form\cite{Heyes2007}:
\begin{eqnarray}
\widetilde{\Phi}(k) = \frac{1}{2\pi^2k}\int_0^{\infty} r \, \Phi(r) \sin(kr)dr \geq 0
\label{Eq:finalCriteria}
\end{eqnarray}
that must be verified for all $k$. Then, applying this criterion to our 
potential (replacing Eq. \ref{Eq:SoftPotential} in Eq. \ref{Eq:finalCriteria}) and 
integrating, an inequality in terms of $w$ is obtained:
\begin{eqnarray}
\widetilde{\Phi}(k) = \frac{\lambda^3 e^{-k^2 \lambda^2} \left(\sqrt{2} (1+w) e^{k^2 \lambda^2/2} - 4 w \right) }{4 \pi ^{3/2}} \geq 0
\label{Eq:FinalConditionPotential}
\end{eqnarray}
where is easy to see that the sign of this expression is given for the 
expression inside the parenthesis:
\begin{eqnarray}
\sqrt{2} (1+w) e^{k^2 \lambda^2/2} - 4 w \geq 0
\label{eq:Inq1}
\end{eqnarray}
as all other terms are positive. Moreover, in Eq. \ref{eq:Inq1} is clear that the 
term $e^{k^2 \lambda^2/2}$ is greater or equal to 1 for all $k$ values, and in 
particular the inequality is verified independently of $w$ if $k \rightarrow 
\infty$. Then and since this inequality should be satisfied for all $k$ values, 
without loss of generality we can take $k=0$, so the final condition is reduced 
to:
\begin{eqnarray}
\sqrt{2} (1+w) - 4 w \geq 0
\label{eq:Inq2}
\end{eqnarray}
which is always satisfied if:
\begin{eqnarray}
w \leq (2^{3/2}-1)^{-1}
\label{eq:Inq3}
\end{eqnarray}
Finally, taking $w$ in the interval $0 \leq w \leq (2^{3/2}-1)^{-1}$ ensures the 
thermodynamic stability for our system.

\begin{acknowledgement}
{We are grateful to Prof. Martin Kr\"{o}ger (ETH Z\"{u}rich) for his help with the $Z1$ algorithm.}
\end{acknowledgement}

%
%
%

\bibliography{Entanglements_in_FSF}

\end{document}